\documentclass{INTERSPEECH2023}

\interspeechcameraready

\title{GenerTTS: Pronunciation Disentanglement for Timbre and Style Generalization in Cross-Lingual Text-to-Speech}
\name{Yahuan Cong, Haoyu Zhang, Haopeng Lin, Shichao Liu, Chunfeng Wang,\newline
Yi Ren, Xiang Yin, Zejun Ma}
\address{ByteDance}
\email{\{congyahuan, zhanghaoyu.aries, linhaopeng, liushichao, 
wangchunfeng, \newline
ren.yi, yinxiang.stephen, mazejun\}@bytedance.com}
\begin{document}

\maketitle

\begin{abstract}
% Cross-lingual timbre and style generalizable text-to-speech (TTS) aims to synthesize speech with a specific reference timbre or style that is never trained in the target language. Existing TTS methods cannot handle it very well due to the following reasons: 1) the timbre and pronunciation are strongly correlated in the training dataset that usually does not contain multi-lingual speech for a specific speaker; 2) the style and pronunciation are mixed since the speech style contains language-agnostic and language-specific parts. To address these challenges, we propose GenerTTS, a cross-lingual TTS which disentangles pronunciation 
% (e.g., speech content and accent) and timbre/style via self-supervised representation bottleneck and mutual information constraint. 
% Specifically, 1) we elaborately design a HuBERT-based information bottleneck to separate and disentangle timbre and pronunciation;
% 2) we introduce a mutual information constraint to the speech style modeling in the text encoder, which minimizes the mutual information between style and language and thus discards the language-specific information in the style embedding. 
% In inference, we can generate speech with any timbre and style reference that is unseen to the target language in training. 
% Experimental results show that GenerTTS outperforms baseline systems in terms of style similarity and pronunciation accuracy and enables cross-lingual timbre and style generalization\footnote{The audio samples are available at https://accountfordemo.github.io/GenerTTS/}.

Cross-lingual timbre and style generalizable text-to-speech (TTS) aims to synthesize speech with a specific reference timbre or style that is never trained in the target language. It encounters the following challenges: 
1) timbre and pronunciation are correlated since multilingual speech of a specific speaker is usually hard to obtain;
2) style and pronunciation are mixed because the speech style contains language-agnostic and language-specific parts. 
To address these challenges, we propose GenerTTS, which mainly includes the following works:
1) we elaborately design a HuBERT-based information bottleneck to disentangle timbre and pronunciation/style;
2) we minimize the mutual information between style and language to discard the language-specific information in the style embedding. 
The experiments indicate that GenerTTS outperforms baseline systems in terms of style similarity and pronunciation accuracy, and enables cross-lingual timbre and style generalization\footnote{The audio samples are available at https://bytecong.github.io/GenerTTS/}.

\end{abstract}

% intro
% 第一段: 扩充第一句Cross-lingual timbre and style generalizable text-to-speech (TTS) aims to synthesize speech with a specific reference timbre or style that is never trained in the target language. 讲应用场景和任务的重要性
% 第二段：扩充第二句Existing TTS methods cannot handle it very well due to the following reasons: xxx，更详细地解释2种correlation和解耦的必要性，以及不解耦会导致什么问题，以及language-agnostic and language-specific parts分别代表什么，以及之前的模型如果不解耦language-specific，会导致什么 (就是中式英语这个例子）
% 第三段：扩充第三句及之后方法：To address these challenges xxx，这里介绍我们的方法 跟前面的2个challange对应起来，每个novelty对应解决一个challenge。infer和train的过程都要介绍一下，以及infer能实现什么事情。 对于hubert部分，要讲清楚我们分析了hubert的bottleneck特性，呼应abs里的elaborately design，
% 第四段：讲实验结果，做了哪些实验和分析，验证了什么事情

\section{Introduction}
Cross-lingual timbre and style generalization text-to-speech (TTS) generates speech with any specific speaker and style that are unseen for the target language in training. This technique is important and useful for several scenarios and applications: 1) training a multi-lingual expressive TTS while the timbre and speech style can not cover all languages in training dataset, especially for some low-resource languages: we can transfer style and timbre learned in rich-resource languages to the target low resource language. 2) Automatic dubbing, which aims to replace all speech contained in a video with that in a different language, matching the original timbre, and rhythm \cite{federico2020speech,cong2022learning}: we can synthesize target language speech with the same style and timbre using very limited training data via cross-lingual timbre and style generalization.
% while there is a growing demand for more varied and expressive synthesized speech with distinct timbres.

The main challenge of cross-lingual timbre and style generalizable TTS is the decoupling of pronunciation, timbre and style from each other. To be specific, 1) timbre and pronunciation are correlated since multilingual speech of a specific speaker is usually hard to obtain in the training dataset; 2) style and pronunciation are usually strongly mixed since the speech style contains language-agnostic and language-specific parts and language is strongly related to pronunciation.

% One way to achieve cross-lingual timbre and style generalization 
One way to address the challenge
is domain adversarial training. 
% For example, 
Zhang et al. \cite{zhang2019learning} applied gradient reversal layer (GRL) and residual encoder to reduce the timbre on text encoder. However, training a network with GRL is known to be unstable and sensitive to the hyper-parameter setting \cite{mun2022disentangled}. Another way is data augmentation: Sun et al. \cite{sun2020building} proposed a system with augmented data generated by voice conversion. However, the data construction process is relatively complicated, and using the generated speech as the ground truth data for the training process will lead to a decrease in the quality of the synthesized speech. 

Recently, many useful speech representations have been proposed, such as Phonetic PosteriorGrams (PPGs) \cite{cao2020code}, ASR bottleneck features (ASR-BNFs) \cite{zhu2022multi,dai2022cloning} and self-supervised learning (SSL)-based features (wav2vec 2.0 \cite{baevski2020wav2vec} and HuBERT \cite{hsu2021hubert}). The most important feature of these representations is that they can 
% easily 
disentangle the
speech into pronunciation, timbre and other components, which can be acted as the information bottleneck. Taking HuBERT representation as an example, we conduct some analyses on HuBERT (see Section \ref{HuBERT-analysis}) and find that it is proficient at preserving style and pronunciation information while removing timbre information with an appropriate channel size and chosen layer.

% Motivated by this observation, we propose a two-step method that effectively separates timbre, style, and pronunciation from each other: 1) disentangle timbre from pronunciation and style, and 2) disentangle style from pronunciation. Since it can generalize timbre and style in cross-lingual TTS, we call it GenerTTS. Our method is based on SSL representation HuBERT and mutual information (MI) minimization constraint. Specifically, 1) to disentangle timbre from style and pronunciation, we apply HuBERT as the bottleneck feature in our TTS model, ensuring pronunciation robustness and speaker similarity in cross-lingual scenes. 2) To disentangle style from pronunciation, based on our self-supervised presentation-based structure, we propose a new style adaptor that models fine-grained style and removes language-specific characteristics by introducing a mutual information minimization constraint. 

Motivated by this observation, we propose a two-step method that effectively separates timbre, style, and pronunciation from each other: 1) to disentangle timbre from style and pronunciation, we apply HuBERT as the bottleneck feature in our TTS model, ensuring pronunciation robustness and speaker similarity in cross-lingual scenes; 2) to disentangle style from pronunciation, based on our self-supervised presentation-based structure, we propose a new style adaptor that models fine-grained style and removes language-specific characteristics by introducing mutual information (MI) minimization constraint. Since it can generalize timbre and style in cross-lingual TTS, we call it GenerTTS. 

% We conduct extensive experiments and the results 
Experimental results
demonstrate that GenerTTS performs well against baseline models for cross-lingual timbre and style generalization TTS in terms of style similarity and pronunciation accuracy. 
% We also conduct some analyses to verify the effectiveness of our proposed components.

\begin{figure*}[!t] 
 \centering
 \vspace{-0.3cm}
  \includegraphics[width=\linewidth]{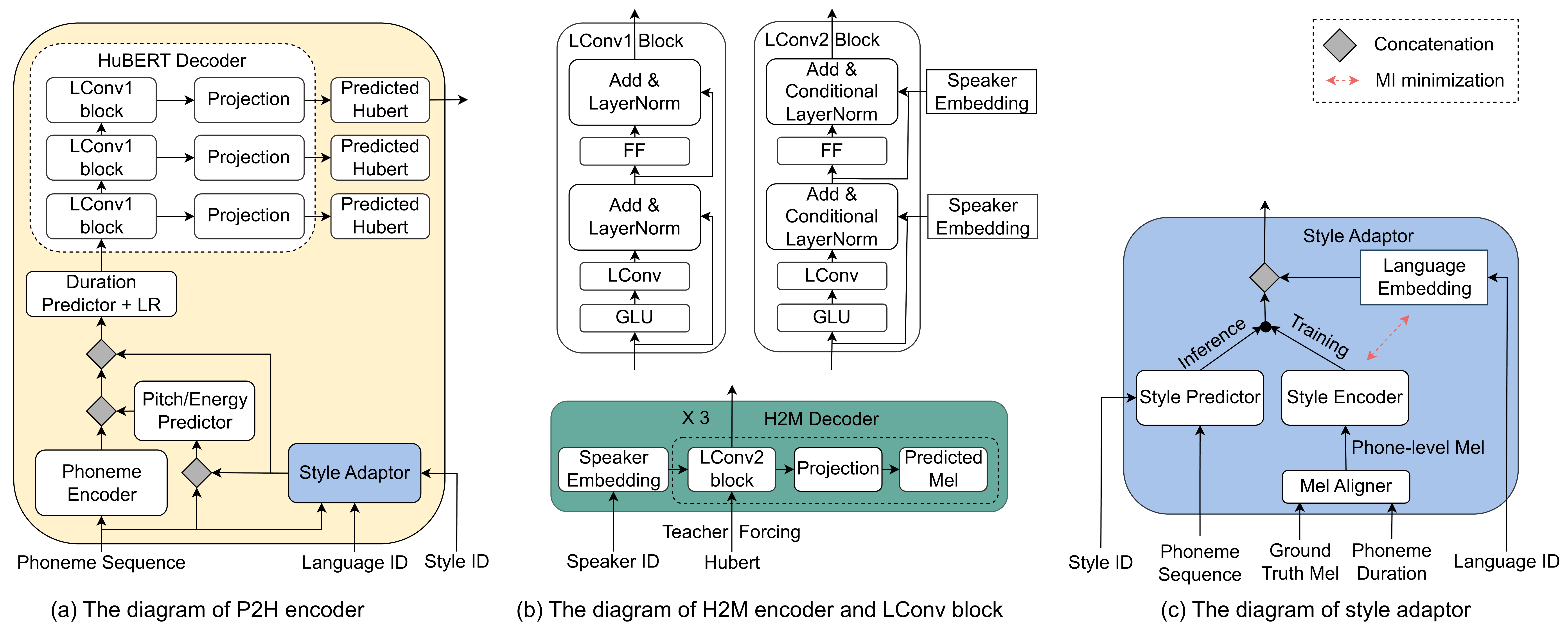}

 \vspace{-0.2cm}
 \caption{The overall architecture for the proposed system.}
 \label{fig: proposed model}
\end{figure*}

\section{Background}

% The generalizable cross-lingual TTS refers to the synthesis of multilingual speech of one speaker even if there is only monolingual data of the speaker. It solves the problem of decoupling timbre and pronunciation to improve timbre similarity and pronunciation accuracy for cross-lingual speaker transfer, and cross-lingual style transfer is not included.
\textbf{Cross-Lingual Style Generalization:} 
% Xin et al. \cite{xin2020cross} propose a system, using gradient reversal and similarity regressor on speaker encoder to reduce language dependency in speaker space. Korte et al. \cite{de2022data} utilize a teacher multi-lingual model to generate teacher-forced data and augmented cross-lingual data to help the student model to alleviate burdens on cross-lingual synthesis. 
The work of Shang et al. \cite{shang2021incorporating} presented a method to deal with the cross-lingual transfer of timbre and style incorporating a fine-grained encoder and gradient reversal modules. However, this method does not disentangle language and style, which will lead to non-native pronunciation for cross-lingual style transfer. For example, the Chinglish phenomenon occurs when the Chinese style is transferred to the English target language.
% Zhang et al. \cite{ratcliffe2022cross} used a Learned Conditional Prior VAE to model Language-independent style information, but accent score dropped for some styles.

\noindent\textbf{Self-Supervised Speech Representation}: Choi et al. \cite{choi2021neural} proposed a neural analysis and synthesis model by using multi-lingual self-supervised learning (SSL) wav2vec 2.0 features as parts of their bottleneck representations, achieving the state-of-the-art in voice conversion and successfully performing multi-lingual voice conversion. 
Du et al. \cite{du2022vqtts} improved the performance of TTS by utilizing a self-supervised VQ acoustic feature instead of traditional mel-spectrogram.
Besides, SSL HuBERT \cite{hsu2021hubert,lakhotia2021generative,polyak2021speech} had better generation results than wav2vec 2.0 and CPC \cite{van2018representation}, and outperformed VQ-VAE \cite{van2017neural} in both speech synthesis and voice conversion. Compared with other supervised bottleneck representations, SSL representation can be trained by plenty of unlabelled speech data.

\section{Proposed Approach}
In this section, we first describe the HuBERT representation. Then we introduce the overall design of the self-supervised representation-based TTS structure proposed in this paper. Finally, we introduce the key techniques to address challenges in cross-lingual style and timbre generalization.

\subsection{Self-supervised acoustic feature: HuBERT}
HuBERT is a self-supervised representation learning model, which 
% can be used for speech recognition and generation tasks and 
shows the potential to disentangle timbres and other features \cite{van2018representation,schneider2019wav2vec,baevski2020wav2vec,hsu2021hubert}. It uses ``acoustic unit discovery system" to generate pseudo labels as the target of iterative training. Besides, a masking strategy similar to BERT \cite{devlin2018bert} is applied in the pretraining to reduce prediction error and help learn representations of long-range temporal relationships. There are three iterations for HuBERT pertaining. K-means on MFCC is the training target for the first iteration, and the output of the trained model can be expected as a better representation than MFCCs. Then, for the second and third iterations, the K-means on the output of the middle layer for the previous iteration are used as the training target of the current iteration. Through such three iterations, a more refined and better continuous representation of pseudo-label can be obtained.

Previous research further discretizes this continuous embedding and applies it to generation tasks. However, discretization leads to the loss of prosodic information, which we expect more to be preserved in Hubert. Therefore, we use continuous embedding and verify that continuous embedding shows great performance in removing timbre information while retaining pronunciation and style. We will show our experiments and analysis in Section \ref{HuBERT-analysis}. Given the aforementioned properties of HuBERT, we are motivated to utilize it as the bottleneck feature in our GenerTTS system to disentangle timbre and pronunciation, as well as timbre and style.

\begin{figure}[h]
 \centering
 \includegraphics[width=0.9\linewidth]{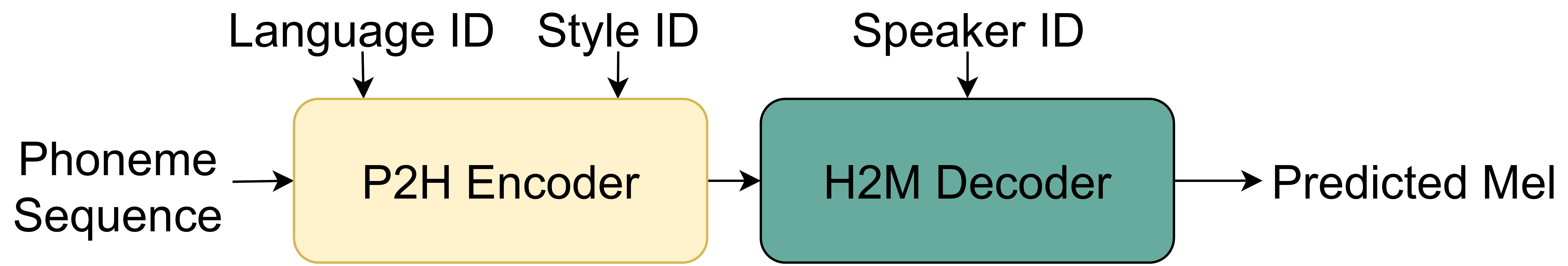}
 \caption{The overall pipeline of our system.}
 \label{fig: Pipilein_TransferableSpeech}
\end{figure}

\subsection{SSL-based TTS system}
Our GenerTTS consists of a P2H (phoneme sequence to HuBERT) encoder and an H2M (HuBERT to Mel-spectrogram) decoder, as shown in Figure \ref{fig: Pipilein_TransferableSpeech}.
P2H as text encoder is used to predict HuBERT embedding from the input phoneme sequence.
H2M is designed for timbre adaptation, which refers to taking HuBERT as input and generating matching Mel-spectrograms according to different speaker embedding conditions.
Waveform is synthesized from predicted Mel-spectrogram by a neural vocoder, which in our experiment is Melgan \cite{kumar2019melgan}. 

\subsubsection{P2H encoder}
Our P2H encoder provides style and pronunciation information. As shown in Figure \ref{fig: proposed model}(a), the structure of P2H consists of a phoneme encoder, a pitch/energy predictor, a duration predictor with length regulator (LR), and a HuBERT decoder. The P2H also includes a style adaptor for modeling fine-grained style, which is described in Section \ref{style_adaptor}.
Our P2H is based on parallel tacotron \cite{elias2021parallel}, which is a highly parallelizable neural TTS model with self-attention and lightweight convolutions. 
In the P2H encoder, a phoneme encoder learns latent representation from phoneme sequence, and then the representation is fed to a duration predictor to predict phoneme duration. And according to the duration information, the LR upsamples the phoneme encoder output to the length of the target frame sequence. Furthermore, similar to the parallel tacotron, an iterative loss is also used for the predicted output in our P2H encoder.

Innovatively, we propose a decoder to predict the target HuBERT embedding from the upsampled phoneme encoder output, instead of directly predicting Mel-spectrogram. Our HuBERT decoder contains three lightweight convolution blocks (LConv1 blocks) with a fully-connected layer. The LConv1 block consists of a gated linear unit (GLU), a lightweight convolution (LConv), a feedforward (FF) layer, and two residual connections with layer normalization as shown in the upper part of Figure \ref{fig: proposed model}(b), which follows \cite{elias2021parallel} \cite{vaswani2017attention}.

In addition, to increase the stability of our proposed model and explicitly model prosody information, pitch predictor and energy predictor are added to our proposed model as in \cite{zhan2021exploring}. Specifically, we extract pitch and energy from speech waveform and map them to phoneme-level features depending on phoneme duration. We take them as conditional inputs in training and use predicted values in inference. Pitch and energy predictors are jointly trained with the proposed model.

\subsubsection{H2M decoder}
Our H2M decoder, shown in the lower part of Figure \ref{fig: proposed model}(b), predicts Mel-spectrogram from HuBERT and supports timbre information for synthesized speech. The network includes a 3-layer LConv2 Block with a fully-connected layer, and the target is 80-dimensional Mel-spectrogram. For high timbre adaptation quality, LConv2 Block is conditioned on speaker embedding by replacing the layer normalization of LConv1 Block with conditional layer normalization, which has been verified efficiency for timbre adaptation in \cite{chen2021adaspeech}. Moreover, teacher forcing is used for H2M training, by making the ground truth HuBERT embedding as input to the H2M decoder during training and using the predicted HuBERT during inference. The iterative loss is also used between the predicted spectrogram and the ground truth Mel-spectrogram.

\subsubsection{Cross-lingual style adaptor}\label{style_adaptor}

As shown in Figure \ref{fig: proposed model}(c), our style adaptor mainly consists of four parts: Mel-spectrogram aligner (Mel aligner), style encoder, style predictor, and language embedding with mutual information constraint.

% In order to separate style and pronunciation, a style encoder is utilized to model fine-grained style. To achieve this, Mel aligner is used to transform the frame-level Mel-spectrogram into a phoneme-level spectrogram according to phoneme duration information and employ the phoneme-level spectrogram as the input of the style encoder. 
% Afterward, the style encoder extracts fine-grained style representation from the input, which is then concatenated with the output of the phoneme encoder. 

We utilize a style encoder to model fine-grained style embedding from phoneme-level spectrogram and concatenate it with the output of the phoneme encoder.
Mel aligner is used to map the frame-level Mel-spectrogram into a phoneme-level spectrogram according to phoneme duration information.
% and employ the phoneme-level spectrogram as the input of the style encoder. 
Since variance features such as pitch and energy are closely related to style, we add style embedding to the pitch and energy predictor. 

For cross-lingual style transfer, the spoken content and even the language of the Mel-spectrogram referenced during inference are inconsistent with the input in the style encoder.
Therefore, a style predictor is utilized to address these inconsistencies. The style predictor adopts text-related information to predict the fine-grained style conditioned on the style ID. In training, the output of the style encoder is used as the prediction target of the style predictor after the stop gradient. For inference, we use the style predictor to predict style depending on the target style ID and input phoneme sequence. The networks of style encoder and style predictor are the same with \cite{shang2021incorporating}. 

In addition, for cross-lingual style transfer, we want the synthesized speech to have a high style similarity while still having the native pronunciation of the target language. In order to improve the pronunciation nativeness for style transfer, we further decouple the language-specific information from the style embedding. 
We model the language embedding from language ID and minimize the MI between language embedding and style embedding. Due to the difficulty of estimating MI in high-dimensional space, we minimize the upper limit of MI measured by the variational contrastive log-ratio upper bound (vCLUB) \cite{cheng2020club}.

\section{Experiments and Results}
\subsection{Experimental setup}
% We evaluate our proposed model using proprietary datasets in four languages, which are Mandarin Chinese (zh-CN), English (en-US), Japanese (ja-JP), and Brazilian Portuguese (pt-BR). 
% zh-CN has a total of 94.6 hours with 4 speakers, en-US has a total of 43.0 hours with 5 speakers, ja-JP has a total of 52.1 hours with 5 speakers, and a total of 69.3 hours with 5 speakers in pt-BR. The sampling rate of those datasets is 24k.
% We evaluate our proposed model using proprietary datasets in two languages, which are Mandarin Chinese (zh-CN) and English (en-US). zh-CN has a total of 94.6 hours with 4 speakers, en-US has a total of 43.0 hours with 5 speakers. 
We evaluate our proposed model using 94.6 hours of Mandarin (zh-CN) from 4 speakers and 43.0 hours of English (en-US) from 5 speakers.
The sampling rate of those datasets is 24k.

For the HuBERT embedding in our experiments, we train HuBERT models using 2000 hours of proprietary datasets in Chinese and English. The sampling rate of this dataset is 16k. Our HuBERT model based on fairseq \cite{ott2019fairseq} consists of 7 CNN layers and 12 layers transformers. Some former researches show that outputs from different layers of the SSL model may contain different information \cite{shah2021all,choi2021neural,hsu2021hubert}. Empirically, we used the continuous embedding output by the 9th layer as the target of P2H of our proposed system due to retaining more style information and pronunciation information.

% The data information is shown in Table \ref{tab: Dataset} and the sampling rate of all datasets is 24k.
% \begin{table}[h]
% \caption{Training dataset breakdown}
% \label{tab:Dataset}
% \centering
% \begin{tabular}{lllll}
% \toprule
% \textbf{Languages} & \textbf{Hours} & \textbf{Speakers} & 
% \textbf{ Utterances}\\
% \midrule
% zh-CN & 94.6 & 4 & 70567 \\
% en-US & 43.0 & 5 & 29730 \\
% ja-JP & 52.1 & 5 & 38439 \\
% pt-BR & 69.3 & 5 & 38512 \\
% \bottomrule
% \end{tabular}
% \end{table}
To validate the performance of our proposed model, two baselines are implemented:
\begin{itemize}
\item \textbf{Para}: Parallel tacotron. The model details follow \cite{elias2021parallel}. And for comparison, we add pitch and energy to the system, which is the same as our proposed system.
\item \textbf{M3}: A multi-speaker multi-style multi-language speech synthesis baseline system \cite{shang2021incorporating}. We reproduce the baseline system based on the basic structure of Para. 
\item \textbf{Ours}: Our proposed cross-lingual speech synthesis system, which is based on HuBERT and style adaptor.
\end{itemize}

\begin{table*}[t]
 \caption{Evaluation score of speaker similarity, style similarity, pronunciation nativeness MOS and ER (CER/WER). W indicates within-lingual and C indicates cross-lingual in the table.}
 \label{tab: Style Similarity}
 \centering
 \begin{tabular}{llllll}
 \toprule
 \textbf{Model} & \textbf{Speaker Similarity } &
 \textbf{Style Similarity} &
 \textbf{Pronunciation Nativeness} &
 \textbf{ER (\%)}\\
 \midrule
 Para & \textbf{0.922} (W:0.967; C:0.877) & 3.43 (W:3.32; C:3.54) & 3.35 & 13.42 (W:4.01; C:22.82) \\
 M3 & 0.920 (W:0.966; C:0.875) & 3.71 (W:3.35; C:4.07) & 3.71 & 12.12 (W:3.96; C:20.28) \\
 \midrule
 Ours & 0.921 (W:0.964; C:0.878) & \textbf{3.99} (W:3.90; C:4.07) & \textbf{4.13} & \textbf{ 9.97 } (W: 4.02; C:15.93) \\
 \midrule
 Ours w/o MI & / & 3.92 (W:3.90; C:3.95) & 3.71 & 12.38 (W:3.58; C:21.19) \\
 Ours w/o Adaptor & / & 3.81 (W:3.57; C:4.06) & 3.63 & 12.54 (W:3.87; C:21.20) \\
 % w/o Adaptor\&HuBERT (=Para) & x & x & x &/ & x \\
 % w/o HuBERT & / & x & x & x & 13.83 (w:4.51; c:23.15) \\
 \bottomrule
 \end{tabular}
\end{table*}

\subsection{Result}
\subsubsection{System evaluation}
For cross-lingual timbre and style generalization, we evaluate the cross-lingual synthetic speech generated by transferring one timbre and two styles from Chinese to English. Both style and timbre are unseen in the target language during training. Moreover, the timbre comes from female voice data, and the two styles are female customer service and male novel narration.
We also evaluate the within-lingual synthetic speech generated by this timbre and these two styles in Chinese. 

For objective evaluation, we utilize cosine similarity to measure speaker similarity, and employ Character Error Rate (CER) for zh-CN and Word Error Rate (WER) for en-US to evaluate pronunciation accuracy.
For subjective evaluation, we conduct a mean opinion score (MOS) experiment on a 5-point scale (5:excellent, 4:good, 3:fair, 2:poor, 1:bad) to evaluate style similarity. Moreover, pronunciation nativeness MOS is evaluated on a 5-point scale, which is only available for cross-lingual.
Table \ref{tab: Style Similarity} shows the experimental results. 

% \begin{table*}[t]
% \caption{Evaluation score of speaker similarity, style similarity, pronunciation nativeness MOS and ER (CER/WER). W indicates within-lingual and C indicates cross-lingual.}
% \label{tab: Style Similarity}
% \centering
% \begin{tabular}{llllll}
% \toprule
% \textbf{Model} & \textbf{Speaker Similarity} &
% \multicolumn{2}{c}{\textbf{Style Similarity}}
% & \multirow {\textbf{Pronunciation Nativeness}}
% & \multirow {\textbf{ER (\%)}} \\
% \cmidrule (l){3-4} 
% & & \textbf{Within-lang} & \textbf{Cross-lang} \\
% \midrule
% Para & \textbf{0.922} (w:0.967; c:0.877) & x & 3.54 & 3.35 & 13.42 (w:4.01; c:22.82) \\
% M3 & 0.920 (w:0.966, c:0.875) & x & \textbf{4.07} & 3.71 & 12.12 (w:3.96; c:20.28) \\
% \midrule
% Ours & 0.921 (w:0.964; c:0.878) & \textbf{3.46} & \textbf{4.07} & \textbf{4.13} & \textbf{ 9.97} (w: 4.02, c:15.93) \\
% \midrule
% Ours w/o MI & / & 3.45 & 3.95 & 3.71 & 12.38 (w:3.58; c:21.19) \\
% Ours w/o Adaptor & / & x & 4.06 & 3.63 & 12.54 (w:3.87; c:21.20) \\
% % w/o Adaptor\&HuBERT (=Para) & x & x & x &/ & x \\
% % w/o HuBERT & / & x & x & x & 13.83 (w:4.51; c:23.15) \\
% \bottomrule
% \end{tabular}
% \end{table*}

Compared with the two baseline systems, our system has obvious advantages in pronunciation accuracy for cross-lingual speech synthesis, which is reflected in the pronunciation of native and ER (CER/ WER). And our system has also shown improvement in style similarity, especially compared with Para.
It should also be noted that the style similarity in cross-lingual is higher than in within-lingual. This is because, in cross-lingual style transfer, the target style and the synthesized speech are in different languages, which makes the evaluators prioritize global style and pay less attention to style details. However, the style MOS criteria are the same for both within-lingual and cross-lingual evaluations.

\subsubsection{Ablation study}
We further conduct an ablation study to validate the components in GenerTTS, which include SSL HuBERT and style adaptor with MI constraint on the linguistic information of style features, as shown in Table \ref{tab: Style Similarity}. 
%\subsubsection{Ablation on style adaptor}
The MI removed system has a sharp decline in pronunciation nativeness MOS and increases in cross-lingual Pronunciation ER. Although MI leads to an increase in pronunciation ER for within-lingual, the results are still basically the same as the baseline model. After removing the style adaptor, the performance deteriorates further, but higher than Para. These results support that the proposed style adaptor and MI are crucial for the disentanglement of language and style, and enable style transfer across languages while maintaining native pronunciation. 

%\subsubsection{Ablation on HuBERT}
In comparison to the Para (ours w/o HuBERT and style adaptor), the system with HuBERT (ours w/o style adaptor) demonstrated higher style similarity and pronunciation nativeness MOS score, as well as lower pronunciation ER in within-lingual and cross-lingual synthesis. This shows that HuBERT has played a significant role in the decoupling of timbre and pronunciation/style.
% as well as the decoupling of timbre and style.
These results demonstrate the effectiveness of each component in our proposed model.

\begin{table}[t]
% \centering
 \caption{HuBERT-based voice conversion, where F indicates female speaker
and M indicates male.}
 \label{tab:hubert}
 \centering
 \begin{tabular}{ccc}
 \toprule
 \textbf{Model} & \textbf{Style Similarity} & \textbf{Cosine Similarity} \\
 \midrule
 M2M & 3.98 & 0.948 \\
 F2M & 3.86 & 0.946 \\
 F2F & 4.06 & 0.970 \\
 M2F & 4.22 & 0.946 \\
 \bottomrule
 \end{tabular}
\end{table}

\subsection{HuBERT analysis} \label{HuBERT-analysis}
\begin{figure}[t]
 \centering
 \vspace{-0.3cm}
 \includegraphics[width=0.8\linewidth]{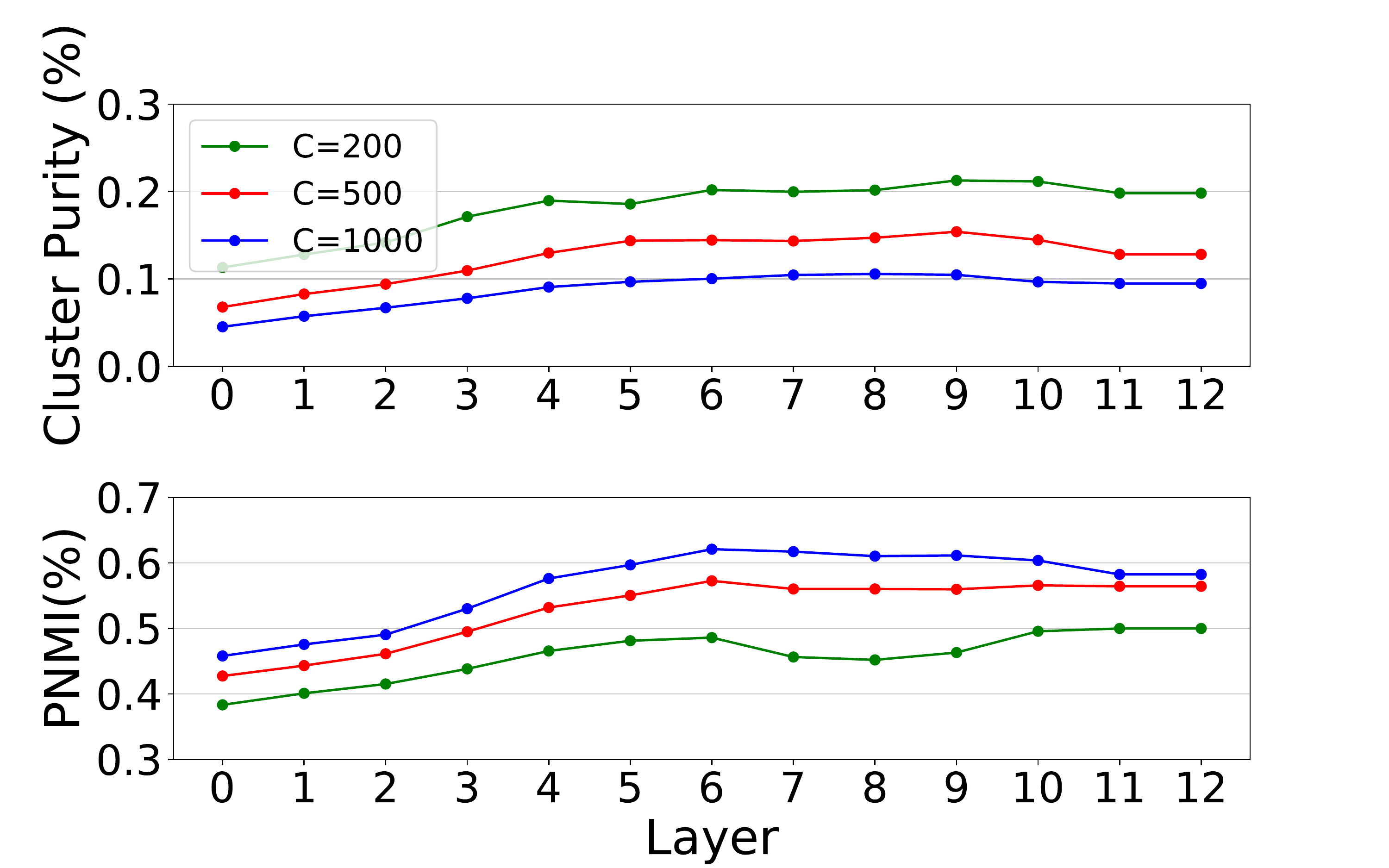}
 \vspace{-0.2cm}
 \caption{Quality of cluster assignments by running k-means on HuBERT embedding from different layers.}
 \label{fig:hubert_layer}
\end{figure}

In this section, we separately validate the pronunciation, timbre, and style information in HuBERT. We first evaluate the pronunciation information of different layers. Same with the original HuBERT paper, we apply k-means features of all 13 layers of the HuBERT model into 200 clusters, 500 clusters, and 1000 clusters. Cluster Purity and Phone-Normalized Mutual Information (PNMI) are also calculated as shown in Figure \ref{fig:hubert_layer}. A higher indicator means the embedding for clustering is more pronunciation-related. Hence we use the embedding from the 9th layer for its relatively higher numbers.

% We then conduct a voice conversion experiment to examine the timbre and style information in HuBERT. 
We then examine the timbre and style information in HuBERT by voice conversion.
We train an independent H2M decoder to serve as voice conversion. HuBERT embeddings are extracted from data of two unseen Chinese speakers and used as input to the voice conversion.

\begin{figure}[t]
 \centering
 \vspace{-0.3cm}
 \includegraphics[width=0.8\linewidth]{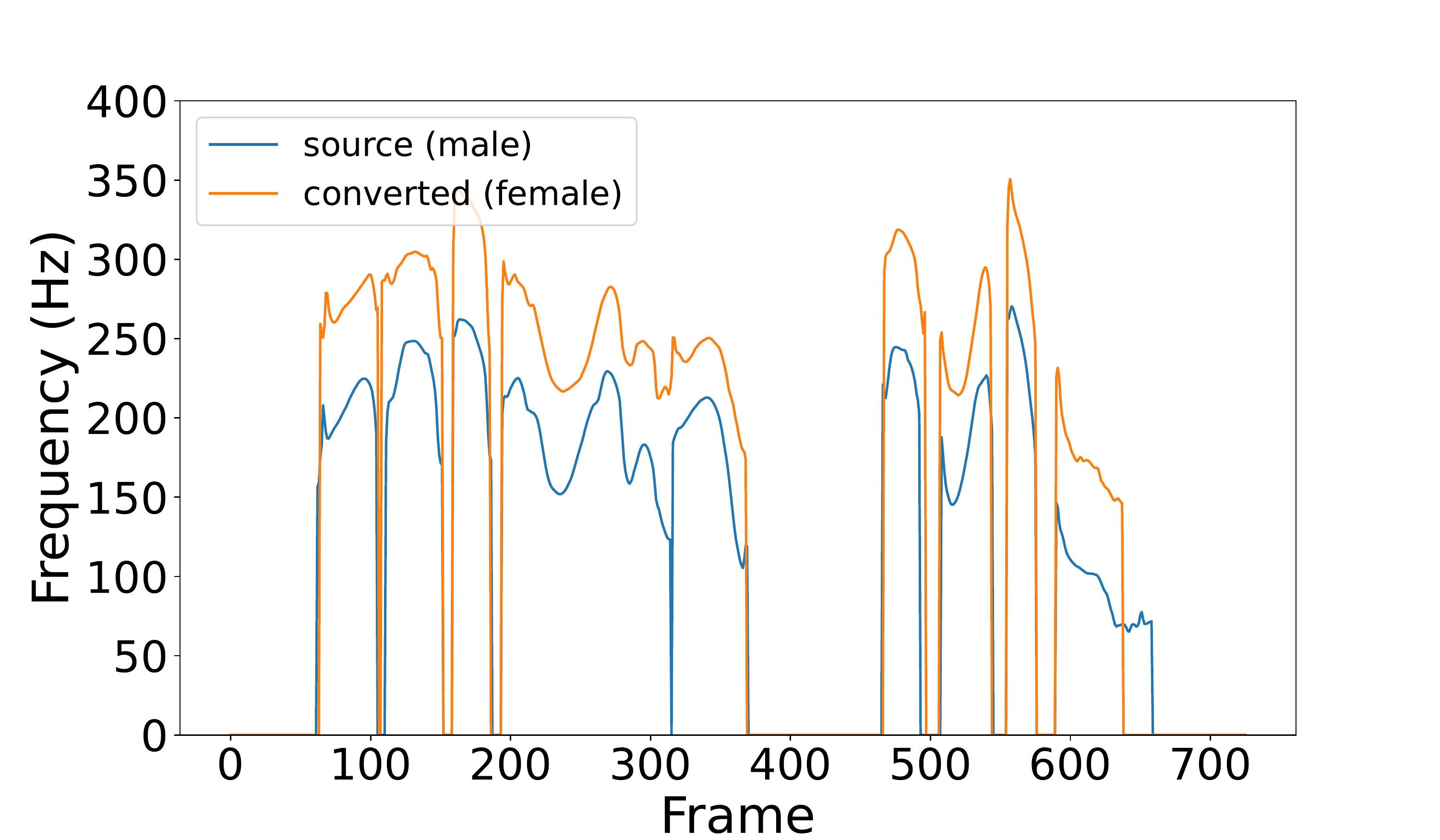}
 \vspace{-0.2cm}
 \caption{The pitch of the source audio and converted audio.}
 \label{fig:vc}
\end{figure}

We calculate the cosine similarity of speaker verification embedding between generated audio and target timbre audio. As shown in Table \ref{tab:hubert}, the H2M decoder can generate audio with high speaker cosine similarity, which indicates that little timbre-related information exists in HuBERT embedding. 
In addition, the higher style similarity means that style information is restored in HuBERT. We also extract F0 from the input and output voices by straight \cite{kawahara2006straight} and they show similar variation trends in Figure \ref{fig:vc}. Those results indicate that HuBERT embedding might be an effective bottleneck feature to remove timbre while retaining other factors, such as style, and pronunciation.

\section{Conclusion}
In this paper, we propose GenerTTS to address cross-lingual timbre and style generalizable speech synthesis with a specific timbre or style that is never trained in the target language.
Experimental results show that GenerTTS outperforms two baseline systems in terms of pronunciation accuracy and style similarity. In addition, ablation experiments show that the HuBERT-based TTS system can improve the pronunciation accuracy of cross-lingual TTS. And our proposed MI can reduce language-specific information in style and improve style similarity and pronunciation nativeness for cross-lingual style transfer. 

\bibliographystyle{IEEEtran}
\bibliography{GenerTTS}

\end{document}